# The Absence of Horizon in Black-Hole Formation

Pei-Ming Ho[1],

*Department of Physics and Center for Theoretical Sciences,*
*Center for Advanced Study in Theoretical Sciences,*
*National Taiwan University, Taipei 106, Taiwan, R.O.C.*

**Abstract**

With the back-reaction of Hawking radiation taken into consideration, the work of Kawai, Matsuo and Yokokura [1] has shown that, under a few assumptions, the collapse of matter does not lead to event horizon nor apparent horizon. In this paper, we relax their assumptions and elaborate on the space-time geometry of a generic collapsing body with spherical symmetry. The geometry outside the collapsing sphere is found to be approximated by the geometry outside the white-hole horizon, hence the collapsing matter remains outside the Schwarzschild radius. As particles in Hawking radiation are created in the vicinity of the collapsing matter, the information loss paradox is alleviated. Assuming that the collapsing body evaporates within finite time, there is no event horizon.

---

[1]e-mail address: pmho@phys.ntu.edu.tw

# 1  Introduction

The information loss paradox [2] has been one of the most interesting and controversial problems in theoretical physics for decades. It is a crucial test of our understanding of black holes when quantum effects are taken into account.

The paradox presents a conflict between the unitarity of quantum mechanics and other basic notions in physics. A major part of the puzzle is how to transfer information from the collapsed matter deep inside the horizon to the Hawking radiation. All proposals of resolution seem to be in conflict with some of our understandings in theoretical physics, e.g. the no-cloning theorem, the monogamy of entanglement, as well as the causality and locality of semiclassical effective theories. In particular, it was recently proven [3] that the information loss can be avoided only if there is order-one correction at the horizon (for instance, a firewall [4]). The possibility we would like to focus on in this paper is that the event horizon never forms as long as the collapsing body completely evaporates within finite time.

For a static black hole, from the viewpoint of a distant observer, it takes an infinite amount of time for her to see an infalling observer to cross the horizon. On the other hand, an infalling observer can pass through the horizon within finite proper time. This is related to the fact that the Eddington advanced time $v$ (which is more relevant to an infalling observer [6]) is related to the Schwarzschild coordinates $t$ and $r$ (which is more relevant to a distant observer) via the relations

$$v = t + r^*, \qquad r^* \equiv r + a \log \left| \frac{r}{a} - 1 \right|. \tag{1}$$

Because $r^* \to -\infty$ when $r \to a$, it is possible that an infinite time $t = \infty$ at $r = a$ corresponds to a finite value of $v$, say $v = v_1$. However, one should be alerted when we take the Schwarzschild radius $a(t)$ to be time-dependent. If the black hole is completely evaporated at $t = t^*$ ($a(t) = 0$ for $t \geq t^*$), it is no longer possible to say that $v_1$ corresponds to $t = \infty$, because $v_1 = t_1 + r^* = t + r$ for $t > t^*$, assuming that eq.(1) still makes sense for time-varying $a(t)$. A more careful analysis [6] shows that if we modify the formula (1) to properly incorporate a time-dependent $a(t)$, it would indeed take infinitely long in his proper time for the infalling observer to pass through the apparent horizon, if $a(t) \to 0$ within finite time. This is in contradiction with the common belief that an infalling observer can pass through the horizon in finite time even if the horizon evaporates completely in finite time. In fact, this common belief typically necessitates a negative energy flux near the horizon.

Classically, the appearance of a horizon also takes infinite time for a distant observer, who sees the collapse of a star getting slower and slower as it gets closer and closer to the point where the horizon emerges. When Hawking radiation is turned on, it is assumed in the conventional model of a black hole [7] that the (apparent) horizon appears in finite time for a distant observer, although it is unclear how Hawking radiation can speed up the formation of horizon. (This is another way to see that a negative energy flux is needed in the conventional model.) Conventionally, it is also believed to be a good approximation to ignore Hawking radiation during the formation of a black hole, and consider Hawking radiation only after the horizon appears. By patching the Penrose diagram of black-hole formation in the absence of Hawking radiation with that of black-hole evaporation, one finds the Penrose diagram in Fig.1 [7], which represents the conventional model of a black hole.

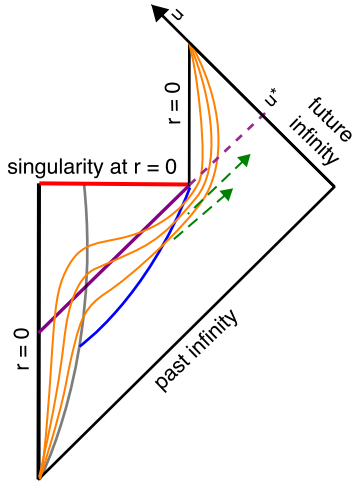

Figure 1: The conventional Penrose diagram for black-hole formation and evaporation: The event horizon (purple line) forms due to a collapsing star (grey curve). The orange curves represent constant $r$ surfaces, which are tangential to outgoing light cones at the Schwarzschild radius (blue curve). The dash green arrows represent Hawking radiation, and the black hole evaporates compleltely at $u = u^*$.

One of the most important features of this diagram is that, the trajectories of the collapsing matter, as well as that of an infalling observer, are terminated at the event horizon at finite Eddington retarded time $u^*$ for a distant observer. Another feature is that the space-time is geodesically complete (Minkowskian) for a distant observer after $u = u^*$, when the black hole completely evaporates. Intriguingly, according to Fig.1, a distant observer would see (through signals at the speed of light) at the same instant ($u = u^*$) the images of all infalling objects disappearing into the event horizon, and at the same moment the black hole evaporates completely with the horizon shrunk to zero size.

There has been suspicion that the conventional model of black holes is incorrect. Some argued that the event horizon should be replaced by a trapping horizon [8]. In the context of string theory, the picture of the fuzzball was proposed to replace the black hole and its horizon [9]. In the collapse of a scalar-field domain wall, there is no horizon either [10]. More recently, through an analysis of the dynamics of the interior of the collapsing star including the back-reaction of Hawking radiation, the same conclusion of no horizon was achieved [11].

Of course, even in the absence of both event and apparent horizon, the appearance of a collapsing star to a distant observer can approximate a real black hole to arbitrary accuracy as the surface of the collapsing sphere is arbitrarily close to the Schwarzschild radius. These objects are sometimes called incipient black holes, black stars, or just black holes. The observational evidence of black holes does not immediately invalidate theories without horizons.

The seminal work of Kawai, Matsuo and Yokokura [1] gives the cleanest and clearest evidence that the conventional Penrose diagram Fig.1 should be discarded. They found that neither apparent horizon (a trapped region) nor event horizon form through the collapse of matter. This was shown elegantly through the consideration of a spherical distribution of matter collapsing at the speed of light. The crucial point is that the formation and evaporation of a black hole are not two separate processes distinct from each other. Hawking radiation should be taken into consideration before the horizon appears, as long as the surface of the collapsing matter is very close to the Schwarzschild radius. [1]

In a previous paper [6], the argument of Ref. [1] for no horizon was simplified, and it was pointed out that the following three assumptions are sufficient to imply that the event horizon of a black hole cannot appear: (i) spherical symmetry, (ii) the Hawking radiation can be approximated by the energy-momentum tensor of outgoing massless dust, and (iii) the collapsing star evaporates within finite time. It is not necessary to assume that the collapsing matter are moving at the speed of light, or that the rate of radiation is given by a certain specific expression.

Following the works of Kawai et al [1, 12, 13] and Ref. [6], we aim at sharpening the arguments to give a comprehensive account of the formation and evaporation of an incipient black hole. We shall also elaborate on related physical and geometrical issues to generalize the discussion as much as possible, still assuming spherical symmetry. We make minimal assumptions about properties of Hawking radiation and depict consistent alternative Penrose diagrams in very general situations.

Our understanding is the following. For a collapsing spherical distribution of matter, the collapse slows down for a distant observer as its radius gets close to the Schwarzschild radius. During the period of time when the radius is very close to the Schwarzschild radius, the geometry of the space-time region outside the collapsing sphere is at a short time scale approximately the same as the geometry outside the horizon of a Schwarzschild solution. Hawking radiation is then expected to appear. We find that, however, *at a large time scale*, the geometry of the space-time region outside the collapsing sphere is not approximated by the geometry outside a black-hole horizon, but by that of a white-hole horizon. (Recall that the maximally extended Schwarzschild solution has both a black-hole and a white-hole horizon.) Hawking radiation continues as long as the radius of the collapsing sphere stays close to the white-hole horizon, until the evaporation is completed.

Even though the Schwarzschild metric is the unique vacuum solution to Einstein's equations with spherical symmetry, it is a degenerate description of both white holes and black holes, related to each other by a time-reversal transformation. We know that the metric outside a collapsing sphere with Hawking radiation must be given by a small deformation of the Schwarzschild metric, but a priori it could be either a deformation of the black-hole geometry or the white-hole geometry. [2] Hence we cannot trust the conventional argument for why it takes only finite proper time for an infalling object to cross the horizon, which assumes that the geometry around the horizon is a deformation of the black hole. Instead, we will show below that the other possibility, that the geometry outside the collapsing matter is a deformation of the white-hole geometry, would provide a consistent model for the collapse of a spherical body including its Hawking radiation. (Of course, the geometry inside the collapsing sphere is *not* that of a white hole.)

---

[1] Some call it "pre-Hawking radiation" or "Hawking-like radiation" since it occurs before or without a horizon.

As a property of the geometry of a white-hole horizon, it is impossible for the radius of the collapsing sphere, or an infalling observer, to cross inside the Schwarzschild radius. Everything stays outside the Schwarzschild radius until it shrinks to zero at the end of the evaporation, and then the space-time is empty and geodesically complete.

If the evaporation is completed within a finite time for a distant observer, as it should according to the usual formulas of Hawking radiation, all infalling time-like or light-like trajectories are geodesically complete because they can be smoothly continued into the empty (geodesically complete) space-time at the time of complete evaporation. The space-time geometry is thus free of event horizon [1,6,13].

If the evaporation takes forever due to a deviation of Hawking radiation from the usual formulas, an event horizon can appear at the infinity of the Eddington retarded time, but it leads to no more information loss than a classical black hole. The information hidden behind the event horizon is not accessible to a distant observer, so that she cannot verify unitarity. Yet the information is not destroyed, just like the information outside the horizon in the de Sitter space-time.

Regardless of whether it is a complete evaporation, Hawking radiation is generated around the surface of the collapsing star when its radius is approximately (but slightly larger than) the Schwarzschild radius [1,6,12,13]. The information of the collapsing matter is transferred into Hawking radiation locally as the matter evaporates. The information loss paradox is therefore reduced to a problem in local quantum field theory, although involving gravity.

As the Hawking radiation can have local interactions with the collapsing matter when it is created near the horizon, its quantum state is entangled with that of the collapsing matter. The decoherence of the radiation when it is separated from the collapsing matter then justifies our use of the energy-momentum tensor for a classical radiation even in the neighborhood of the collapsing sphere.

The plan of the paper is as follows. In Sec.2 we review the KMY model with more emphasis on how it provides a consistent picture of the black-hole formation and evaporation, in contrast with the inconsistency of the conventional model. We explain in Sec.3 that a horizon cannot form within finite Eddington retarded time because the geometry outside the collapsing sphere resembles the geometry outside a white-hole horizon. Hawking radiation appears as the surface of the collapsing sphere stays close to the Schwarzschild radius. While a consistent picture of Hawking radiation can be given (See Sec.4), we also consider the possibility that Hawking radiation deviates significantly from the usual formulas (e.g. at a small scale due to new physics) so that an event horizon can exist beyond infinite Eddington retarded time. Then we generalize the KMY model by more general metrics, corresponding to more general processes of collapse and Hawking radiation, in Sec.5 and Sec.6. Finally we summarize and comment on remaining problems in Sec.7.

---

[2] The black-hole has outgoing light-like horizon, and the white-hole an ingoing light-like horizon.

## 2  KMY Model Revisited

In this section we revisit the Kawai-Matsuo-Yokokura (KMY) model [1] for black-hole formation and evaporation. Based on Refs. [1, 6, 12, 13], we will illuminate the geometrical aspects of the model, focusing on the issues relevant to the information loss paradox.

The KMY model describes a spherically symmetric distribution of null dust collapsing at the speed of light, with no assumption about the density distribution. Intuitively, if a horizon can ever appear in a certain collapsing process, it should also be able to appear when everything is collapsing at the highest speed – the speed of light. The conventional expectation is that the whole sphere of dust should eventually fall inside the horizon. But this turns out to be a false belief. The key point of the KMY model is that the evaporation and formation of a black hole take place simultaneously, with the evaporation prohibiting the formation to be completed. [1, 6, 13]

In this paper, we are not concerned with the singularity at the center of the black hole, so we do not have to consider high energy corrections such as loop contributions in quantum gravity. The gravitational theory is purely classical (Einstein's theory). The only quantum effect is the inclusion of the energy-momentum tensor for Hawking radiation, which has a quantum origin.

Let us recall that, even in the conventional model of the black hole, in terms of the Eddington retarded time $u$, Hawking radiation appears *before* the event horizon, so that $a(u)$ decreases with time even when there is still no horizon. The fact that neither event nor apparent horizons is a necessary condition for Hawking radiation has been well established in the literature [14]. It turns out to be crucial that we turn on Hawking radiation before there is a horizon, and even when there will never be a horizon, in order for the theory of a black hole to be self-consistent.

For simplicity, Hawking radiation is assumed to be composed only of null dust and to respect spherical symmetry in this model. (We will consider more general Hawking radiation later in this paper.) For a sufficiently large black hole, the radiation of massive particles is negligible, and the radiation modes with non-zero angular momentum should be sub-leading. Hence we expect this assumption about Hawking radiation not to change conclusions about qualitative features of the black hole.

We will assume that the collapsing sphere has a well-defined outer surface, and focus on the spacetime outside the collapsing matter. We will also assume that the energy-momentum tensor outside the surface of the matter sphere is given only by an outgoing light-like energy flux representing the Hawking radiation. (It can also include other radiation due to the coupling between the collapsing matter and massless fields, but we will focus on the Hawking radiation below for simplicity.) For an arbitrary time-dependence of Hawking radiation, its most general asymptotically flat solution to Einstein's equation is given by the outgoing Vaidya metric: [3]

$$ds^2 = -\left(1 - \frac{a(u)}{r}\right)du^2 - 2dudr + r^2 d\Omega^2 \qquad (r \geq R(u)). \tag{2}$$

Here $R(u)$ denotes the outer radius of the collapsing sphere, and the metric (2) is assumed to be valid only for $r \geq R(u)$. The Schwarzschild radius $a(u)$ is related to the Bondi mass $M(u)$ of the collapsing star by the relation

$$a(u) = 2M(u). \tag{3}$$

(Generalizations of $a(u)$ to $a(u,r)$ will be considered below in Sec.5.)

The light-like coordinate $u$ is called the Eddington retarded time. The mass $M(u)$ is supposed to decrease with time since Hawking radiation takes away energy from the matter. This metric is valid outside the collapsing sphere for any density distribution (as long as it is spherically symmetric) with the same total Bondi mass $M(u)$.

In their semi-classical approximation [1], the Einstein tensor gives, through the Einstein equation $G_{\mu\nu} = 8\pi G \langle T_{\mu\nu} \rangle$, the expectation value of the energy-momentum tensor

$$\langle T_{\mu\nu} \rangle = -\delta^u_\mu \delta^u_\nu \frac{1}{8\pi G} \frac{\dot{a}(u)}{r^2}. \tag{4}$$

Note that the Einstein equation implies that the energy-momentum tensor is conserved, i.e. $\nabla_\mu \langle T^{\mu\nu} \rangle = 0$. Eq.(4) is the most general form of the energy-momentum tensor for an outgoing flux of null dust with spherical symmetry.

Some of the readers may feel uncomfortable with eq.(4). In the conventional model, the Hawking radiation can be treated as classical radiation only at a distance far away from the black hole, and that there should be a negative energy flux near the apparent horizon for the sake of energy conservation. This opinion is a result of the assumption that the geometry outside the collapsing matter is a small deformation of the black-hole geometry — in the sense that an object can fall inside the horizon within finite proper time — and this assumption has led to not only the presence of a negative energy flux, but also the serious theoretical conflict called information loss paradox. On the other hand, we shall simply assume that there is nothing except the outgoing radiation (4). We shall examine the logical consequences of this assumption, and compare it with the conventional model. Of course, our assumption will be justified only if the following questions can be answered:

*Why can eq.(4) describe Hawking radiation in the region close to the horizon?* (5)

---
[3] Sometimes the *ingoing* Vaidya metric is used to describe the interior of the collapsing star in the formation of a black hole from null dust, when the effect of Hawking radiation is ignored.

In fact, we will not only answer this question (at the end of this section), but also show that the information loss paradox is resolved in the KMY model [1].

Let us now examine the consequences of the metric (2). We are concerned with infalling light-like trajectories in radial directions ($d\Omega = 0$), in particular the outer surface of the shell $R(u)$, as well as outgoing null geodesics for the massless particles in Hawking radiation. Due to the continuity of the metric, the trajectory of $R(u)$ is determined by the metric (2). Hence the metric allows us to check whether the outer radius of the matter sphere will enter a horizon. If there will be a horizon, one would expect that eventually the whole sphere collapsing at the speed of light will fall inside the horizon.

According to the metric (2), light-like trajectories must satisfy either of the following two equations:

$$\left(1 - \frac{a(u)}{r}\right) du + 2 dr = 0, \tag{6}$$

$$du = 0. \tag{7}$$

For $r > a(u)$, the infalling light-like curves (e.g. $R(u)$) satisfy the first equation, and the outgoing ones (e.g. massless particles in Hawking radiation) the second. For $r < a(u)$, the roles played by these two equations are interchanged. In order to claim that there is no horizon, one will have to show that the metric (2) is geodesically complete for all infalling causal trajectories.

In the conventional model, a black hole is expected to evaporate completely within finite Schwarzschild time $t$ for a distant observer. Since the usual formula for Hawking radiation in the conventional model is a good approximation of that in the KMY model [1], a distant observer should observe the disappearance of the black hole within finite $u$. [4] We therefore assume here that the black hole evaporates completely at $u = u^*$ for some finite $u^* < \infty$. (We will consider the case of $u^* = \infty$ in Sec.4.4.) That is,

$$a(u) = 0 \quad \text{for} \quad u \geq u^*, \tag{9}$$

and after $u^*$ the spacetime turns into the Minkowski space according to (2).

Now we show that all ingoing null trajectories are geodesically complete. First we note that the trajectory of the Schwarzschild radius $r = a(u)$ is space-like [6], independent of the details of Hawking radiation, as long as $\dot{a} < 0$ because

$$ds^2 = -2\dot{a} du^2 > 0 \tag{10}$$

for $r = a(u)$. That is, as long as the Bondi mass decreases with time (due to Hawking radiation), the Schwarzschild radius is shrinking at a speed faster than light. It is impossible

---

[4] The Eddington retarded time $u$ is related to the Schwarzschild time $t$ by

$$u \simeq t - r \tag{8}$$

at large $r$. A finite change in $t$ is mapped to a finite change in $u$ at large $r$.

for any light-like or time-like ingoing geodesics to cross it from the outside to the inside. In fact, the Schwarzschild radius of the metric (2) is a deformation of the white-hole horizon, as we will explain below in Sec.3.

Recall that the metric (2) is valid only for $r \geq R(u)$. Hence $r = a(u)$ does not really correspond to any real curve in space-time unless $R(u) \leq a(u)$. Since $r = a(u)$ defines a superluminal infalling trajectory, it is impossible for $R(u)$ to get smaller than $a(u)$ if $R(u_0) > a(u_0)$ at the initial time $u_0$, and thus the Schwarzschild radius $r = a(u)$ can only exist via analytic continuation, not in real space-time.

For a generic, monotonically decreasing $a(u)$, with $\dot{a}(u) < 0$ for all $u \leq u^*$, the (fictitious) Schwarzschild radius shrinks at a speed faster than light, and thus over time it keeps a finite difference in its value of $r$ from all ingoing light-like geodesics, until the moment when the evaporation is complete (See Fig. 3). [5] As a result, the curve B in Fig.2, the Penrose diagram for the conventional model, cannot exist, because it demands $R(u^*) = a(u^*) = 0$. More importantly, the curve A in Fig.2 cannot exist either. At $u^*$, every ingoing null trajectories continues from a finite value of $r > 0$ into the empty (Minkowski) space, and can thus be extended to infinity in its affine parameter. In other words, the metric (2) is geodesically complete for all ingoing causal geodesics, in contrast with the existence of geodesically incomplete curves in the conventional model for the space-time visible to a distant observer. This has also been demonstrated via numerical simulations [1, 6] (See Fig.3). The Penrose diagram for the region $r \geq R(u)$ should thus look the same as that for a region of the Minkowski space as Fig.4(a).

The interior region of the collapsing sphere can be approximated by infinitely many infinitesimally thin shells of null dust separated by infinitesimal empty regions [1, 13]. As outer shells cannot affect inner shells due to spherical symmetry, a metric of the same form as (2) applies to every empty region between two neighboring spherical thin shells. The same argument then applies to the outer surface of each thin shell, and we conclude that there is no horizon inside the collapsing sphere either. The interior of the sphere is thus described by the Penrose diagram Fig.4(b), also without horizon. The complete Penrose diagram for the collapsing sphere should thus be obtained by patching up Figs.4(a) with (b), to be given by Fig.4(c) [1]. Fig.4(c) is essentially the same as the Penrose diagram of a Minkowski space.

In the absence of any horizon, the Hawking radiation is created in the neighborhood of the surface of the collapsing sphere. Local interactions would allow the information of the collapsing matter to be transferred to the Hawking radiation. The information loss paradox is resolved up to details involving ordinary local quantum field theories.

Let us now answer the question posted as eq.(5). First, eq.(4) is obtained as the expectation value of the energy-momentum operator for the vacuum state defined in Ref. [1]. Second, in the KMY model, there is never a period of time when the collapsing matter is well separated from the place where Hawking radiation is originated, as the collapsing mat-ter never falls inside any horizon. Particles in the Hawking radiation are always allowed to interact locally with the collapsing matter, so that the quantum state of the radiation are entangled with the collapsing matter. This entanglement thus leads to the decoherence of the radiation when it is separated from the collapsing matter, justifying the treatment of the Hawking radiation as classical radiation.

---

[5] For this we have to assume that $\dot{a}(u^*) < 0$. The possibility $\dot{a}(u^*) = 0$ will be considered in Sec.4.3, and the conventional Penrose diagram is still incorrect.

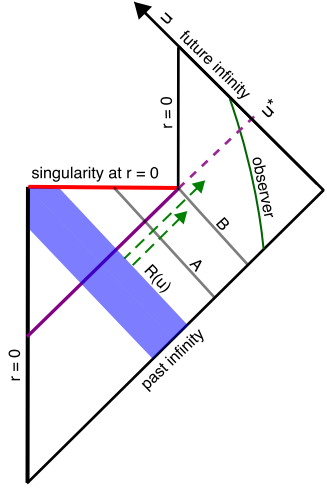

Figure 2: The conventional model of black-hole formation and evaporation for a sphere (started as a shell) of null dust (blue strip) collapsing at the speed of light: Line A is an incomplete null trajectory in the space-time visible to a distant observer, terminated at the event horizon (purple line). Line B is a null trajectory that reaches the origin $r = 0$ at exactly the moment $u = u^*$ when the evaporation completes.

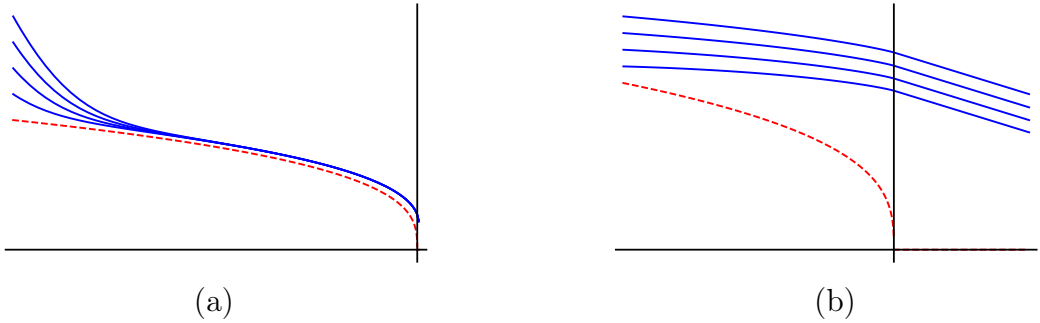

Figure 3: In the KMY model, ingoing null geodesics cannot enter the Schwarzschild radius because the latter shrinks at a speed faster than light. Fig. (a) shows that, initially, when the ingoing massless particles are still far away, they move "quickly" (in terms of the coordinates) towards the origin. But they "slow down" as they get closer to the Schwarzschild radius. Fig. (b) focuses on the behavior of the null trajectories during a short period of time around $u = u^*$ As the Schwarzschild radius goes faster than light, it is at a finite difference in $r$ from other null trajectories, which are continued into flat spacetime after the Schwarzschild radius vanishes. The plots are made assuming that the Hawking temperature is proportional to $1/a$.

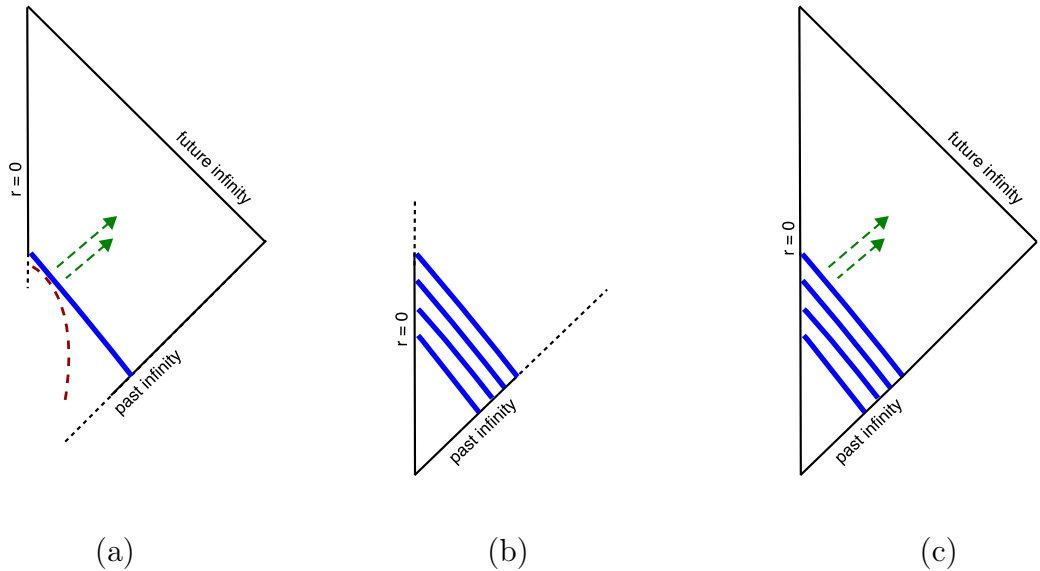

(a) (b) (c)

Figure 4: (a) Penrose diagram for the region outside the collapsing sphere. The dash red curve represents the Schwarzschild radius of the metric (2). The dash green arrows represent Hawking radiation. (b) Penrose diagram inside the collapsing sphere. (c) The complete Penrose diagram for the collapsing sphere obtained by combining the previous two diagrams.

## 3 White-Hole Horizon

According to the metric (2), the tangent vectors on the curves of constant $r$ are null-like at the Schwarzschild radius $r = a(u)$, time-like for $r > a(u)$ and space-like for $r < a(u)$. The surface $r = a(u)$ at the Schwarzschild radius thus resembles an apparent horizon. This does not imply that particles within the region $r < a(u)$ are trapped inside. They are actually "trapped outside", that is, it is impossible for a particle to cross inside the Schwarzschild radius from the outside. In the region $r < a(u)$, both light cone directions (6) and (7) are outgoing. On the other hand, all outgoing null curves can reach $r = \infty$, regardless of whether it is originated from inside the Schwarzschild radius. (If there exists a black-hole horizon in the collapse, it is impossible to see it in the coordinate system of the outgoing Vaidya metric. The absence of black-hole horizon has to be shown through the completeness of the metric, as we did in Sec.2.)

In the classical limit when $a(u)$ is a constant, the space-time metric should reduce to the Schwarzschild solution. Recall that the maximally extended Schwarzschild solution has two horizons: the black-hole horizon and the white-hole horizon. Apparently, if we take the limit of zero Hawking radiation ($\dot{a} \to 0$), the Schwarzschild radius in the KMY model must coincide with the white-hole horizon.

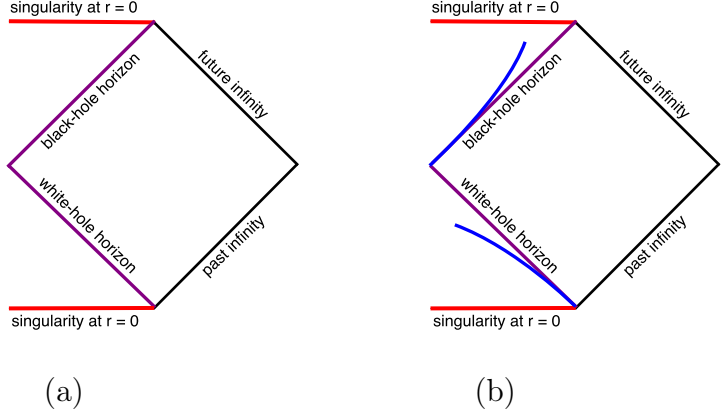

(a)          (b)

Figure 5: Half of the Penrose diagram for the maximally extended Schwarzschild space-time is depicted in Fig. (a). Both horizons (purple lines) are light-like. In Fig. (b) we sketch schematic curves (blue curves) obtained as deformations of the two event horizons by decreasing $r$ (leftward) when increasing $t$ (upward). The black-hole horizon is deformed to be time-like, and the white-hole horizon space-like.

Very naively, it is of course less counter-intuitive to say that radiation comes out of a white hole than a black hole. Another intuitive way to argue for the fact that the Schwarzschild radius corresponds to the white-hole horizon in the classical limit is the following. The Hawking radiation is expected to decrease the Schwarzschild radius, since the size of the Schwarzschild radius should be smaller for a smaller mass in the collapsing matter. If we identify the Schwarzschild radius as a deformation of the event horizon of a black hole, it would be time-like when it shrinks (See Fig.5). On the other hand, if we identify the Schwarzschild radius with a deformation of the even horizon of a white hole, it would be space-like, as we saw in eq.(10).

When the collapsing sphere is extremely close to the Schwarzschild radius, locally the geometry outside the sphere is indistinguishable from the region outside the event horizon for a massive static black hole. For an astronomical collapsing star, the Schwarzschild radius $a(u)$ is almost constant over an extremely long period of time. Hawking radiation is thus expected to appear in the vicinity of the horizon, regardless of whether it is the black-hole horizon or the white-hole horizon.

This understanding helps us resolve a conceptual problem. According to (6), the radius of the outer surface of the collapsing matter approaches to the Schwarzschild radius by the formula

$$R(u) \simeq a(u) - 2a(u)\dot{a}(u) + \cdots \qquad (11)$$

when $R(u)$ is extremely close to $a(u)$. The correction term $-2a(u)\dot{a}(u)$ is always positive so that $R(u) > a(u)$ at all times, but some readers may be worried because this term is of sub-Planckian scale for a large black hole. Can one trust a semi-classical calculation involving Planck-scale physics? Is it possible that quantum mechanically the matter can tunnel inside the horizon and a black hole horizon starts to exist after that?

The puzzle is resolved as we realize that the Schwarzschild radius is that of a (decaying) white hole. Even if the collapsing matter somehow falls inside the horizon for some unknown reason, outgoing light rays can still reach out of the Schwarzschild radius, and the matter shell is forever visible for a distant observer. A black-hole horizon can only exist outside the coordinate patch of $(u, r)$. The superficial sub-Planckian scale in the correction (11) is of a similar nature as the arbitrarily small separation between a particle in constant acceleration and the asymptotic light cone it approaches to.

Incidentally, the formula (11) gives a cutoff of the horizon by

$$\Delta r \simeq -2a(u)\dot{a}(u). \tag{12}$$

This is the shortest difference in coordinate $r$ from the horizon anything can get to. In the conventional model of black holes, the horizon can be reached by an infalling observer, for whom the energy density of the Hawking radiation is blue-shifted by the factor $(1 - a/r)^{-1}$ which diverges at the horizon. This infinite energy flux of Hawking radiation at the horizon leads to the puzzle about the so-called "firewall" [4]. In the KMY model, the maximal blue-shift factor is

$$\left(1 - \frac{a}{r}\right)^{-1} \simeq \left(\frac{-2a(u)\dot{a}(u)}{R(u)}\right)^{-1} \simeq \frac{1}{2|\dot{a}(u)|}. \tag{13}$$

As the energy flux of Hawking radiation is proportional to $|\dot{a}(u)|$, there is no divergence in energy flux anymore.

To be more explicit, consider an infalling trajectory with a normalized time-like tangent vector

$$\hat{n}^\mu = (\hat{n}^u, \hat{n}^r, 0, 0), \qquad \hat{n}_\mu \hat{n}^\mu = -1, \tag{14}$$

where

$$\hat{n}^u = \frac{e^\zeta}{\sqrt{1 - a/r}}, \qquad \hat{n}^r = -\sqrt{1 - a/r}\sinh\zeta \tag{15}$$

for any $\zeta > 0$. The energy-momentum tensor (4) of Hawking radiation in his reference frame is

$$T_{uu}\hat{n}^u\hat{n}^u = -\frac{1}{8\pi G}\frac{\dot{a}}{r^2}\frac{e^{2\zeta}}{1 - \frac{a}{r}} \simeq \frac{1}{16\pi G}\frac{e^{2\zeta}}{a^2}. \tag{16}$$

For a generic value of $\zeta$, the energy momentum tensor is extremely small for a Schwarzschild radius $a$ of astronomical size.

# 4 Hawking Radiation

## 4.1 Hawking Radiation for Collapsing Sphere

There is a potential problem about the Hawking radiation in the KMY model. While our arguments are so far applicable to an arbitrary radial distribution of energy for the collapsing sphere, we have not shown that the energy flux of Hawking radiation is independent of the energy distribution. Intuitively, when we approximate the matter sphere by many thin matter shells, with each matter shell just outside its Schwarzschild radius defined by the energy enclosed within that shell, we expect each thin shell to create its Hawking radiation, and generically the total Hawking radiation of the whole sphere would depend on the distribution of energy. This would make the radiation of a collapsing sphere different from a black hole. Interestingly, it turns out that the total Hawking radiation is approximately independent of the energy distribution at a late stage when each thin shell is approaching to its Schwarzschild radius according to (11) [1].

The energy flux of Hawking radiation for a collapsing spherical thin shell is given by [1,13] [6]

$$J = \frac{N\hbar}{8\pi}\{u, U\}, \qquad (18)$$

in the eikonal approximation, where $N$ is the number of species of massless free scalars and $\{u, U\}$ is the Schwarzian derivative

$$\{u, U\} \equiv \frac{\ddot{U}^2}{\dot{U}^2} - \frac{2\dddot{U}}{3\dot{U}}. \qquad (19)$$

The coordinates $u$ and $U$ are respectively the Eddington retarded time outside and inside the shell. But there is no assumption about the trajectory of the shell. This formula can be obtained by comparing the vacuum expectation value of the energy-momentum tensor on the two sides of the shell [1].

A generic spherical distribution of collapsing null dust can be approximated by a large number of thin shells separated by very thin layers of empty space. Using $u_n$ as the Eddington retarded time coordinate in the empty layer between the $n$-th and $(n+1)$-th shell (counted from the origin towards the outside), we generalize the formula above to

$$J_n = \frac{N\hbar}{8\pi}\{u_n, u_{n-1}\} \qquad (20)$$

for the energy flux of Hawking radiation generated by the $n$-th collapsing shell. Assuming that there are $L$ layers, we have $u_L = u$ and $u_0 = U$, where $u$ is the Eddington retarded time outside the whole sphere in the metric (2) and $U$ is the light-cone coordinate of the empty space at the origin.

---

[6] The energy flux $J$ is defined such that the Schwarzschild radius changes with time according to

$$\frac{da}{du} = -2GJ. \qquad (17)$$

Note that the energy flux of each shell is red-shifted when it passes through the next shell, by the red-shift factor $(du_n/du_{n+1})^2$. The total energy flux of the first two inner-most shells can be expressed as

$$J_2 + \left(\frac{du_1}{du_2}\right)^2 J_1 = \frac{N\hbar}{8\pi}\{u_2, U\}. \tag{21}$$

Formally, this is the same formula for Hawking radiation as if all the Bondi mass of the two shells is concentrated on the second shell. (This is only formally true, while the functional dependence of $U$ on $u_2$ is different.) This is due to an identity of the Schwarzian derivative

$$\{u_2, u_1\} + \left(\frac{du_1}{du_2}\right)^2 \{u_1, U\} = \{u_2, U\}. \tag{22}$$

Interestingly, using this identity iteratively, one can show that, independent of the distribution of energy over the radius of the collapsing sphere, the energy flux of the Hawking radiation is always given by (18).

Assuming that the approximation (11) applies to all layers in the KMY model, and that the configuration depends only on the Schwarzschild radius of each layer, the energy flux is approximately given by [1]

$$J \simeq \frac{N\hbar}{96\pi}\left(\frac{1}{a^2} + \frac{4}{a^2}\frac{da}{du}\right). \tag{23}$$

While the second term is much smaller than the first, one obtains the usual formula for Hawking radiation

$$\dot{a} \propto 1/a^2 \tag{24}$$

(when only massless particles are considered). This formula implies that the black hole evaporates completely in finite time. [7] More precisely, for $N \gg 1$, eq.(23) implies that $\dot{a} \simeq -1/4$ for $a \ll \sqrt{\frac{N\hbar G}{12\pi}}$. It still leads to complete evaporation in finite time. But we should note that the Hawking radiation formula is subject to modifications as the validity of (23) is not without assumption.

## 4.2 Assumptions About Hawking Radiation

We have shown through the KMY model that the Hawking radiation before the appearance of horizon prevents the horizon from appearing. In our argument, we have assumed that the energy flux of the Hawking radiation is given by the usual formula (24), so that the collapsing body is completely evaporated within finite time. However, it is not totally clear that eq.(24) is valid for all possible collapsing processes, when we consider a generic process different from the KMY model. For example, for a collapsing sphere carrying nonzero charge associated with a gauge symmetry, Hawking radiation is expected to stop at the extremal bound $M \geq Q$. Complete evaporation is not always achievable. There are also several reasons to suspect that Hawking radiation in the new picture can have more complicated behaviors.

---

[7] A single infinitely thin shell cannot evaporate completely in finite time [1]. But a shell, however thin, must spread out to a finite thickness over time, due to the difference in metrics on the two sides of the shell. An infinitely thin shell is hence unphysical.

First, while the horizon in the conventional model of black holes is usually assumed to be in the vacuum state, the region from which Hawking radiation is originated in the KMY Model is not empty, but in the vicinity of the collapsing sphere. For a free field under the influence of nothing but the space-time geometry, its contribution to Hawking radiation is indeed approximated by the usual formulas for a black hole of the same mass. [1, 13]. However, if the interaction between the particles in Hawking radiation and the collapsing sphere cannot be ignored, it is possible that the formulas of Hawking radiation can be significantly modified.

Second, for a charge associated with a global symmetry that is carried by particles of mass $m$, the ADM mass of a collapsing body composed of $N$ of these particles is smaller than $Nm$ due to the gravitational potential. It is thus impossible to completely evaporate all the charges to the spatial infinity without violating charge conservation or energy conservation [13]. [8]

In the new picture without horizon, it is possible that everything involved in a complete evaporation happens at an energy scale lower than the Planck scale [13], and it should be possible to explain everything in a low energy effective field theory. Although the issues mentioned above are irrelevant to the KMY model which is restricted to massless free particles in both the collapsing matter and Hawking radiation, they remind us that there are questions about Hawking radiation that remain to be understood for more general collapsing processes. We should therefore try to make as few assumptions about Hawking radiation as possible in this paper, and consider all possible patterns of Hawking radiation.

Let us recall that the only assumptions we made about Hawking radiation in Sec.2 were the following:

1. There is nothing but outgoing null dust in spherically symmetric distributions.

2. $a(u) = 0$ for all $u \geq u^*$.

3. $\dot{a}(u) < 0$ for all $u \leq u^*$.

We will relax the assumption 1 in Sec.5 and 6, and consider violation of the assumption 2 in Sec.4.4. The relaxation of the last assumption to

3'. $\dot{a}(u) \leq 0$ for all $u \leq u^*$,

allowing $\dot{a}(u^*) = 0$, will be considered in Sec.4.3.

It makes little difference to the space-time causal structure whether $\dot{a}$ is allowed to vanish temporarily before $u^*$, because even for constant $a$, it takes infinite change in $u$ for an infalling particle to cross inside the white-hole horizon.

---

[8] In fact, the baryon number is such an example [13].

Let us first briefly review the usual formulas of Hawking radiation. The conventional formula for Hawking temperature in 4 dimensions is

$$T_H = \frac{C}{a}, \qquad (25)$$

for some constant $C$. Assuming that the radiation is dominated by massless particles, the energy flux density is proportional to $T_H^4$. Since the area of the horizon is $4\pi a^2$, the total energy flux is proportional to $a^{-2}$, so that we obtain eq.(24):

$$\dot{a} = -\frac{C'}{a^2} \qquad (26)$$

for some positive constant $C'$. This leads to the time dependence of $a(u)$ as

$$a(u) \propto (u^* - u)^{1/3}, \qquad (27)$$

and the evaporation is always completed within finite time.

## 4.3 Asymptotic Behavior of the Schwarzschild Radius

In this subsection, we will consider the case $\dot{a}(u^*) = 0$ which was dismissed in Sec.2 because it is incompatible with the usual formula of Hawking radiation. Consider the possibility that the Hawking radiation formula is changed due to new physics, or due to the interactions between the collapsing sphere and the radiation. Let us assume that at the late stage of the evaporation, the Schwarzschild radius $a(u)$ is given by the asymptotic expression

$$a(u) \simeq C''(u^* - u)^n, \qquad (28)$$

for a parameter $n > 0$ and a coefficient $C'' > 0$. For $n \leq 1$, $\dot{a}(u^*) < 0$, in accordance with the assumption 3 mentioned above. For $n > 1$, however, $\dot{a}(u^*) = 0$, and the assumption 3 is violated.

For $n < 1$, one can easily see that the the qualitative behavior of ingoing null geodesics are always the same as described above, shown in Fig.3, namely, they reach the origin at a finite time after $u^*$. For $n \geq 1$, however, the trajectory of $R(u)$ may exhibit qualitatively different behaviors. For $n = 1$ with $C'' \leq 1/8$, and for $n > 1$ for generic $C''$, a class of solutions of $R(u)$ has $R(u^*) = 0$. [9] All null geodesics within a certain range of the Schwarzschild

---

[9] For $n = 1$, $C'' \leq 1/8$, there is a continuous set of solutions of $R$ to (6) with the asymptotic expression

$$R(u) \simeq \alpha(u^* - u) + \beta(u^* - u)^n, \qquad (29)$$

where $\alpha = \frac{1}{4}(1 - \sqrt{1 - 8C''})$, $n = \frac{1}{2\alpha} - 1$ and arbitrary (but small) $\beta$. For $n > 1$, we have

$$R(u) \simeq [C''(u^* - u)^n + \cdots] + A e^{-\frac{(u^* - u)^{-(n-1)}}{2(n-1)}}, \qquad (30)$$

where the first part is any solution to (6), and a correction term with an arbitrary (but small) parameter $A$ is allowed.

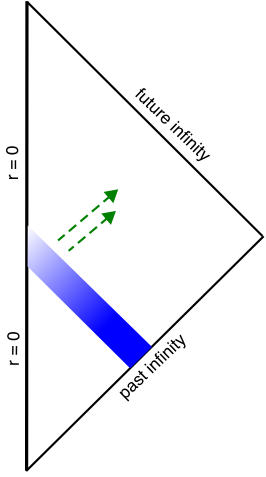

Figure 6: Penrose diagram of the KMY model [1].

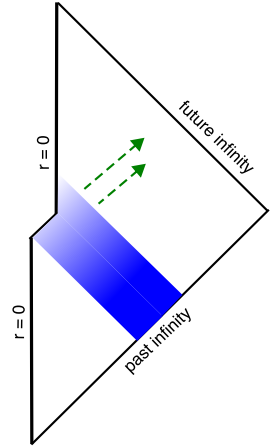

Figure 7: Penrose diagram for the case $a(u) \propto (u^* - u)^n$ with $n > 1$.

radius reach the origin at the same instant (in Eddington retarded time) when the black hole is completely evaporated. This is still inconsistent with the Penrose diagram Fig.2 of the conventional model because there the null trajectory $A$ is terminated at some $r > 0$, rather than $r = 0$. On the other hand, the Penrose diagram of this case is not exactly like that of the KMY model (Fig. 6), but should be given by Fig.7. Although this is different from the Penrose diagram of Minkowski space, the whole spacetime is visible to a distant observer and there is no horizon.

## 4.4 Permanent Black Holes

If the formulas of Hawking radiation is significantly changed, so that it is too weak to evaporate the collapsing sphere in finite time, a logical possibility, although not a necessity, is that there exists a black-hole event horizon, just like a classical black hole in the absence of Hawking radiation. (See [15] for a review on theories of black-hole remnants, including semi-classical ones.)

Let us consider this possibility by starting with the classical black hole without Hawking radiation. The ingoing null trajectories are no longer geodesically complete in the coordinate system of $(u, r)$. The black-hole horizon resides at $u = \infty$ (Fig.8).

As Hawking radiation is gradually turned on (as a parameter that we tune to change the theory), the remnant of radiation gets smaller, and the event horizon is smaller. For Hawking radiation large enough, the collapsing matter evaporates completely in finite time, and there is no black-hole horizon. The effect of gradually turning on Hawking radiation is shown in Fig.9. Strictly speaking, the Penrose diagram only shows the causal structure, and so Figs.9 (a), (b) and (c) are equivalent. But heuristically, we imagine that the effect of turning on Hawking radiation changes the space-time structure from (a) to (d), by reducing the size of the event horizon.

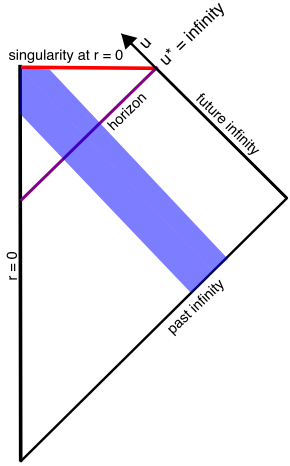

Figure 8: Penrose diagram for a classical black hole without Hawking radiation.

Notice that if we tune the parameter $C$ in the formula of Hawking temperature (25), the collapsing sphere always evaporate completely in finite time as long as $C > 0$. There is complete evaporation regardless of how small the parameter $C$ is. What we imagine to tune is a modification of the usual formula of Hawking radiation, which may be due to new physics at a shorter length scale when the collapsing body is very small (but large enough for semi-classical gravity to apply). It can also happen when the collapsing process is more complicated than what we have depicted above.

For instance, consider a modification to the equation for decay rate (26) by a higher derivative term

$$\dot{a}(u) = -\frac{C}{a^2(u)} - b\,\ddot{a}(u), \tag{31}$$

for some constant $b > 0$. When the Schwarzschild radius is large, the solution of $a(u)$ resembles the usual formula for Hawking radiation. But the second term in (31) dominates over the first when $a$ is small, and its value approaches to a constant as $u \to \infty$, with the Hawking radiation dies off exponentially at large $u$. In this case, there can be a remnant of black hole of mass of order $(bC)^{1/3}$.

Notice that, even in the case of the Penrose diagram Fig.8, the information is still not lost in the Hawking radiation, as the Hawking radiation is generated in the neighborhood of the collapsing matter. The information remained inside the event horizon will never come out and it will be impossible for a distant observer to verify unitarity. Instead of a violation of unitarity, we have a situation in which unitarity cannot be fully verified. This is just like the information of everything in the universe that is outside our causal past.

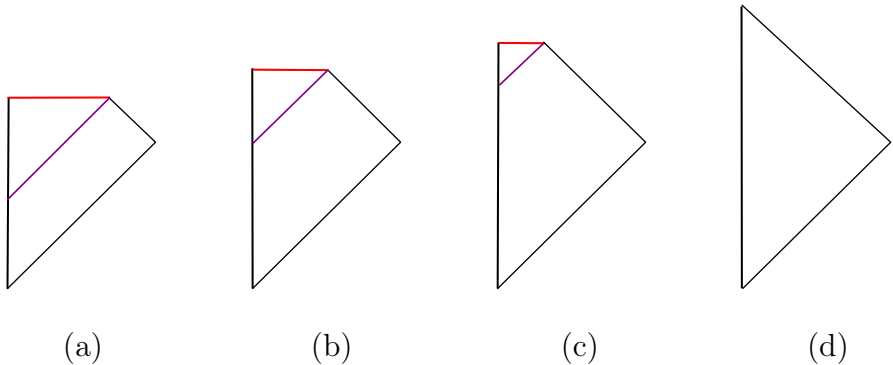

(a) (b) (c) (d)

Figure 9: Penrose diagrams for different theories, in which the effect of Hawking radiation changes from incomplete evaporation to complete evaporation.

## 5 Generalized KMY Model

In this section, we consider a model more general than the KMY model. Spherical symmetry is still assumed, but the assumption on the content of Hawking radiation is relaxed.

We assume that the energy-momentum tensor of the Hawking radiation outside the collapsing sphere is an arbitrary combination of Type-I and Type-II matter fields. Type-I matter fields are those whose energy-momentum tensors can be diagonalized, and Type-II matter fields' energy-momentum tensors have double null eigenvectors. The most general spherically symmetric line element is then parametrized by two functions $\psi(u,r)$ and $a(u,r)$ as [16]

$$ds^2 = -e^{2\psi(u,r)}\left(1 - \frac{a(u,r)}{r}\right)du^2 - 2e^{\psi(u,r)}dudr + r^2 d\Omega^2. \tag{32}$$

The Einstein equation implies that

$$\langle T_{uu}\rangle = \frac{1}{8\pi G}\frac{1}{r^2}e^{\psi}\left[-\dot{a} + e^{\psi}\left(1 - \frac{a}{r}\right)a'\right], \tag{33}$$

$$\langle T_{ur}\rangle = \frac{1}{8\pi G}\frac{1}{r^2}e^{\psi}a', \tag{34}$$

$$\langle T_{rr}\rangle = \frac{1}{8\pi G}\frac{2}{r}\psi', \tag{35}$$

and $\langle T_{\theta\theta}\rangle$ and $\langle T_{\phi\phi}\rangle$ are omitted here ($\sin^2\theta\langle T_{\theta\theta}\rangle = \langle T_{\phi\phi}\rangle$). Primes represent derivatives with respect to $r$, and dots derivatives with respect to $u$.

For instance, if $\psi = 0$ and $\dot{a} \neq 0$, both the weak and strong energy conditions imply

$$\dot{a}(u,r) < 0, \qquad a'(u,r) \geq 0, \qquad a''(u,r) \leq 0. \tag{36}$$

The dominant energy condition further implies that $a'(u,r) \geq r|a''(u,r)|/2$. All of these energy conditions demand that $a(u,r)$ for any given $r$ can only decrease with time. However, we will not assume that any of these energy conditions are satisfied, as it is believed that Hawking radiation may violate some of the energy conditions.

Only asymptotically flat configurations are considered in this paper, so we shall assume that

$$\psi(u,r) \to \text{constant} \quad \text{and} \quad a(u,r)/r \to 0 \quad \text{as} \quad r \to \infty. \tag{37}$$

Hence $u$ can be identified with the retarded time $t - r$ in Minkowski space as $r \to \infty$. The spacetime patch of $(u, r)$ is geodesically complete at least at sufficiently large $r$. If the incipient black hole evaporates away in a finite time for a distant observer, we have $a(u, r) \simeq 0 \ \forall u > u^*$ with some finite $u^*$.

For $r > a(u, r)$, the curves along constant $u$ are outgoing null geodesics that extend to $r = \infty$. If there is an event horizon in spacetime, this patch of the coordinate system $(u, r)$ should be geodesically incomplete, with ingoing null geodesics (analogues of line A in Fig.2) terminated at finite $u$. Conversely, one would conclude that there is no horizon if all infalling geodesics are complete in this coordinate system $(u, r)$.

We use this metric to describe the region $r \geq R(u)$, where $R(u)$ is the outer radius of the collapsing matter. With the metric analytically continued into the region $r < R(u)$, we expect that this metric admits the definition of a Schwarzschild radius $a_0(u)$ satisfying

$$a(u, a_0(u)) = a_0(u) \quad \forall u. \tag{38}$$

If there is no divergence in $a(u, r)$, one can assume that $[1 - a(u, r)/r] > 0$ (i.e. $a(u, r) < r$) for $r > a_0(u)$. Otherwise there is another Schwarzschild radius at a larger radius, and we shall just refer to that solution as $a_0(u)$ instead. In other words, we assume that $a_0(u)$ is the largest solution to eq.(38). [10]

The tangent vectors of constant $r$ are light-like at $r = a_0(u)$, time-like at $r > a_0(u)$ and space-like at $r < a_0(u)$ (but with $r$ larger than the next solution to eq.(38)). This tells us that it is the apparent horizon of a white hole at $r = a_0(u)$.

The story is then very similar to the KMY model. If $\dot{a}_0(u^*) < 0$, all ingoing trajectories are smoothly continued into the geometry described by the metric (32). Assuming that $e^\psi$ does not diverge anywhere, the metric is regular for $u > u^*$ for all $r$, and all causal infalling geodesics are complete in the positive $u$ direction. The Penrose diagram would look like Fig.6. If, on the other hand, $\dot{a}_0(u^*) = 0$, those infalling trajectories which follow $a_0(u)$ sufficiently closely may reach the origin at the same time $u^*$ when $a_0$ vanishes. the Penrose diagram would then be given by Fig.7.

If, however, the factor $e^\psi$ diverges at some point $u = u^*(r)$, or equivalently $r = r^*(u)$, the ingoing trajectories are terminated at $u^*(r)$ depending on their coordinate in $r$. If $u^*(r)$ is independent of $r$, $u^*$ has to be $\infty$ since the space-time is asymptotically flat, with $u$ being approximately $t - r$ at large $r$ in terms of the Schwarzschild coordinates $t, r$.

---

[10] Strictly speaking, it is possible that for a period of time, say, from $u_1$ to $u_2$, there exist a pair of solutions $a_1(u), a_2(u)$ larger than $a_0(u)$, appearing simultaneously at $u_1$ at the same point $r = a_1(u_1) = a_2(u_1)$, and annihilated at $u_2$ with $a_1(u_2) = a_2(u_2)$. They form a closed surface outside the surface of the collapsing matter. This is not a trapped region, but more like an apparent horizon of a white hole (an "untrapped surface"), and thus their existence does not affect the arguments below.

The conventional Penrose diagram Fig.1 suggests that $r^*(u)$ is smaller for larger $u$, with $r^*(u) = 0$ for $u \geq u^*$. Due to the divergence of the factor $e^\psi$ at $r = r^*(u)$, the metric is well-defined only for $r > r^*(u)$. If $r^*(u) \leq a_0(u)$ for all $u$, the divergence of $e^\psi$ at $r = r^*(u)$ is irrelevant since it is hidden behind the white-hole horizon. If $r^*(u) > a_0(u)$, on the other hand, an ingoing trajectory $\tilde{r}(u)$ is terminated at $u^*$ when $r^*(u^*) = \tilde{r}(u^*)$.

A divergence in $e^\psi$ at a finite $r = r^*(u)$ implies a dramatic deviation from the collapsing sphere of null dust. It remains a question whether such a geometry can be realized by physical sources. Of course, any Penrose diagram, physical or not, can be realized by some metric, and plugging that metric into Einstein's equation gives an energy-momentum tensor. One always needs some guidelines to know which diagram is physical. What we have shown above is that the arguments for the KMY model is robust against regular deviations from the outgoing Vaidya metric (2).

## 6  Incipient Black Hole in Schwarzschild Coordinates

For those who suspect that there is something peculiar in using the Eddington retarded time, we consider here an incipient black hole in the Schwarzschild coordinates $(t, r)$. It will be shown to have the same qualitative features as the generalized outgoing Vaidya metric.

The most general metric for spherically symmetric static configurations can be written in the form

$$ds^2 = -e^{-\Phi(r)}\left(1 - \frac{a(r)}{r}\right)dt^2 + \frac{dr^2}{1 - \frac{a(r)}{r}} + r^2 d\Omega^2. \tag{39}$$

If we turn on the time dependence of the two functional parameters, we get

$$\begin{aligned}
ds^2 &= -e^{-\Phi(t,r)}\left(1 - \frac{a(t,r)}{r}\right)dt^2 + \frac{dr^2}{1 - \frac{a(t,r)}{r}} + r^2 d\Omega^2 \\
&\equiv -f(t,r)dt^2 + h(t,r)dr^2 + r^2 d\Omega^2.
\end{aligned} \tag{40}$$

We assume that $a(t,r)/r \to 0$ and $\Phi(t,r) \to 0$ so that the space-time is asymptotically flat.

The Einstein equation gives

$$\langle T_{tt} \rangle = \frac{1}{8\pi G}\frac{e^{-\Phi}}{r^2}\left(1 - \frac{a}{r}\right)\frac{d}{dr}a(t,r), \tag{41}$$

$$\langle T_{tr} \rangle = \frac{1}{8\pi G}\frac{1}{r^2\left(1 - \frac{a}{r}\right)}\dot{a}(t,r), \tag{42}$$

$$\langle T_{rr} \rangle = -\frac{1}{8\pi G}\frac{1}{r^2\left(1 - \frac{a}{r}\right)}\frac{d}{dr}a(t,r) - \frac{1}{r}\frac{d}{dr}\Phi(t,r), \tag{43}$$

and $\langle T_{\theta\theta} \rangle$ and $\langle T_{\phi\phi} \rangle$ are omitted here.

Let us define a quantity $\rho$ that is intuitively the energy density by

$$\rho(t,r) \equiv T_{tt}(t,r)f^{-1}(t,r), \tag{44}$$

and then the energy inside a sphere of radius $r$ can be defined by

$$M(t,r) \equiv \int_0^r dr' \, 4\pi r'^2 \rho(t,r'). \tag{45}$$

It turns out that $a(t,r) = 2M(t,r)$.

Let $R(t)$ denote the radius of the outer surface of the collapsing matter. The metric above is assumed to be valid only for the region $r \geq R(t)$. Nevertheless, we can analytically continue $f(t,r)$ and $h(t,r)$ into the region $r < R(t)$.

The Schwarzschild radius is by definition located where $f(t,r)$ and $h^{-1}(t,r)$ vanishes:

$$1 - \frac{a(t,r)}{r} = 0. \tag{46}$$

We refer to the largest solution [11] to this equation $a_0(t)$, so that

$$a_0(t) = a(t, a_0(t)). \tag{47}$$

Assuming asymptotic flatness, i.e., $h(t,r)$ approaches to 1 as $r \to \infty$, we have

$$h(t,r) > 0 \quad \text{for} \quad r > a_0(t). \tag{48}$$

We shall also assume that $e^{\Phi(t,r)}$ is nonzero and finite for $r > a_0(t,r)$.

Null geodesics are described by the equations

$$\frac{dr}{dt} = \pm e^{-\Phi(t,r)/2} \left(1 - \frac{a(t,r)}{r}\right), \tag{49}$$

where the plus (minus) sign is for outgoing (ingoing) null geodesics for $r > a_0(u)$.

At the point just outside the Schwarzschild radius, $r = a_0(t) + \epsilon$ ($\epsilon > 0$, $\epsilon \to 0^+$),

$$ds^2 = h(t, r = a_0(t)^+) \dot{a}_0^2 dt^2 > 0, \tag{50}$$

so it is always space-like. [12] For $\dot{a}_0 < 0$, a particle originated at a point with $r > a_0(t) + \epsilon$ can never catch up with the trajectory $r = a_0(t) + \epsilon$ as it is shrinking at a superluminal speed. For the same reason, the surface of the collapsing sphere remains outside the Schwarzschild radius.

At the end of the evaporation, $a_0(t)$ goes to zero while all time-like and null-like ingoing geodesics are geodesically complete as they are extended into the empty (Minkowski) space in the absence of collapsing matter.

Here we consider in more detail a special case of the metric (40), which is analyzed in detail in Ref. [13]:

$$ds^2 = -\left(1 - \frac{a(t)}{r}\right) dt^2 + \frac{dr^2}{1 - \frac{a(t)}{r}} + r^2 d\Omega^2. \tag{51}$$

---

[11] As the case of the generalized outgoing Vaidya metric (32), it is possible to have closed trapping surfaces outside $a_0$. See footnote 10.

[12] The trajectory is space-like also when $\dot{a} > 0$. This describes the case when energy flux flows continuously into the black hole and the black-hole horizon expands at a superluminal speed.

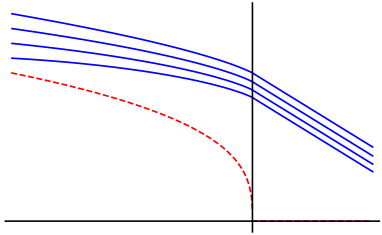

Figure 10: The horizontal axis labels $t$ and vertical axis $r$. For the metric (51), ingoing null trajectories (blue lines) originated outside the Schwarzschild radius (red dash line) are continued into the Minkowski space after the Schwarzschild radius shrinks to zero.

Assuming that the Schwarzschild radius changes with time according to the usual formulas of Hawking radiation for a black hole of radius $a(t)$ with only massless particles in the radiation, we show ingoing null trajectories in Fig.10. Just like the case of the outgoing Vaidya metric shown in Fig.3, the Schwarzschild radius $a(t)$ reaches zero with all ingoing null trajectories at finite $r$, and the ingoing null trajectories can be extended into the Minkowski space after $a(t) = 0$. As the Minkowski space is geodesically complete, none of the ingoing null trajectories are geodesically incomplete from the viewpoint of a distant observer. This implies that there is no horizon, in contrast with the conventional model, as well as many other models of black holes.

## 7 Comments

The two major assumptions made in our arguments are

1. Spherical symmetry.

2. The validity of the semi-classical Einstein equation

$$G_{\mu\nu} = \kappa \langle T_{\mu\nu} \rangle \tag{52}$$

   outside the collapsing sphere, where $\langle T_{\mu\nu} \rangle$ is the expectation value of the energy-momentum tensor of Hawking radiation.

We have considered the most general metrics (32) and (40) with spherical symmetry for both complete and incomplete evaporations. Our consideration covers a very wide range, if not all possibilities, of models of collapsing matter and Hawking radiation. We hope that the physical intuition provided in this work will convince the reader that the assumption of spherical symmetry is not so much an essential ingredient of the argument as merely a technical convenience. On the other hand, there may be important features (which are not directly related to the global causal structure of space-time) such as chaos, that are missed in the spherically symmetric approximation.

The semi-classical approximation implies, as a result of the Bianchi identity, that $\langle T_{\mu\nu} \rangle$ satisfies the conservation law $\nabla_\lambda \langle T_{\mu\nu} \rangle = 0$. The expectation value of $T_{\mu\nu}$ is thus non-zero at the horizon whenever there is Hawking radiation. This may seem odd from the viewpoint of the conventional model, for which the horizon is empty and Hawking radiation is purely a quantum state (at least in the early stage of the black hole evaporation). In the KMY model, however, particles in the radiation interact with particles in the collapsing body, when the former are created around the surface of the latter. The entanglement between the Hawking radiation and the collapsing matter has the effect of a measurement on the Hawking radiation. The energy-momentum tensor of the Hawking radiation can therefore be treated as a classical quantity.

We point out in the above that the space-time geometry outside the collapsing sphere resembles the geometry outside the apparent horizon of a decaying white hole. The ultimate fate of the collapsing matter may be either complete evaporation within finite time (Fig.6 or Fig.7, depending on the property of Hawking radiation at the late stage of evaporation) or incomplete evaporation (Fig.8, if there is an event horizon, or Fig.6 if not).

In all cases, the information loss paradox is alleviated because the particles in Hawking radiation detected by a distant observer are always originated from the surface of the collapsing body, with which they have direct local interactions. In the case of complete evaporation, for both Penrose diagrams Fig.6 and Fig.7, the full space-time is in the causal past of the trajectory of a distant observer. In the case of incomplete evaporation, the causal past of the trajectory of a distant observer misses the region behind the event horizon at $u = \infty$ (See Fig.8). However this black hole is composed only of the remnant of the evaporation, whose information will not be transferred through Hawking radiation to a distant observer. The information is not lost but only inaccessible to the distant observer.

All the information to be carried by the particles in Hawking radiation for the sake of unitarity can be acquired through local interactions with the collapsing matter. It is an ordinary scattering process. The initial state is composed of the vacuum and the collapsing matter. It is scattered into the final states of particles in Hawking radiation and the collapsing matter (with less energy). The unitarity of quantum mechanics is preserved as long as the scattering is calculated in a consistent quantum field theory. (See, e.g. [17].) Nevertheless, to completely resolve the information loss paradox, there are still many open questions to answer: Can we extend our conclusions to cases without spherical symmetry? What is the fate of the collapse of a charged star? Will charges for global symmetries be conserved? Can one calculate the S-matrix of the above-mentioned scattering process in general? We leave these important questions for future study. Following [1, 6, 12, 13], this paper provides a self-consistent description of black holes that has no obvious conflicts with local quantum field theories, and can hence be a good starting point to answer these questions.

# Acknowledgement

The author would like to thank Hikaru Kawai for sharing his original ideas on which this work is based, and Carlos Cardona, Jiunn-Wei Chen, Yi-Chun Chin, Keisuke Izumi, Hsien-chung

Kao, Samir Mathur, Yen Chin Ong, Shu-Jung Yang and Dong-han Yeom for discussions, and in particular Chih-Hung Wu for his encouragement and extensive discussions. The work is supported in part by the Ministry of Science and Technology, R.O.C. and by National Taiwan University.